\documentclass[conference,10pt]{IEEEtran}
\usepackage{amsfonts}
\usepackage{mathrsfs}
\usepackage{amssymb,amsmath}

\usepackage{algorithm}
\usepackage{algorithmic}
\usepackage{amsmath,amssymb,epsfig,subfigure}
\usepackage{graphicx}
\usepackage{theorem}
\usepackage{array,color}
\usepackage[compress]{cite}
\usepackage{bm}
\usepackage{verbatim}

\newtheorem{definition}{Definition}

\newtheorem{proposition}{Proposition}
\theoremheaderfont{\normalfont\bfseries}

\ifCLASSINFOpdf
\else
\fi
\begin{document}
%
\title{Low Complexity List Successive Cancellation Decoding of Polar Codes}

\author{\IEEEauthorblockN{Congzhe Cao, Zesong Fei}
\IEEEauthorblockA{School of Information and Electronics\\
Beijing Institute of Technology\\
Beijing, China\\
Email: {2220120185, feizesong}@bit.edu.cn}
\and
\IEEEauthorblockN{Jinhong Yuan}
\IEEEauthorblockA{School of Electrical Engineering \\and Telecommunications\\
University of New South Wales\\
Sydney, Australia\\
Email: J. Yuan@unsw.edu.au}
\and
\IEEEauthorblockN{Jingming Kuang}
\IEEEauthorblockA{School of Information and Electronics\\
Beijing Institute of Technology\\
Beijing, China\\
Email: jmkuang@bit.edu.cn}}
\maketitle

\begin{abstract}
We propose a low complexity list successive cancellation (LCLSC) decoding algorithm to reduce complexity of traditional list successive cancellation (LSC) decoding of polar codes while trying to maintain the LSC decoding performance at the same time. By defining two thresholds, namely ``likelihood ratio (LR) threshold" and ``Bhattacharyya parameter threshold", we classify the reliability of each received information bit and the quality of each bit channel. Based on this classification, we implement successive cancellation (SC) decoding instead of LSC decoding when the information bits  from ``bad" subchannels are received reliably and further attempt to skip LSC decoding for the rest information bits in order to achieve a lower complexity compared to full LSC decoding. Simulation results show that the complexity of LCLSC decoding is much lower than LSC decoding and can be close to that of SC decoding, especially in low code rate regions.
\end{abstract}


%
\IEEEpeerreviewmaketitle

\section{Introduction}
Polar codes are capacity-achieving codes for the class of binary-input discrete memoryless channels (B-DMCs)\cite{Arikan}. In \cite{Arikan}, successive cancellation (SC) decoding is used to recover information bits. Later in \cite{BP}, belief propagation (BP) decoding is employed to achieve better performance in BECs, but in general B-DMCs the specific schedule of the individual messages for BP decoding is a problem. Linear programming (LP) decoding is introduced afterwards without any schedule, but it does not work for other channels except BECs \cite{LP}. Recently, list successive cancellation (LSC) decoding of polar codes is developed and shows significant performance improvement compared with SC decoding. However, much higher decoding complexity is observed for LSC \cite{Ido} \cite{Chen}. Thus, to find a decoding algorithm with both good frame error rate (FER) performance and low complexity is an open interest. In this paper, we propose a low complexity list successive cancellation (LCLSC) decoding algorithm to significantly reduce the complexity of LSC decoding while trying to maintain its FER performance.


\section{Polar Codes and List Successive Cancellation Decoding}

In \cite{Arikan}, polar codes are introduced to achieve the capacity of B-DMCs by exploiting the channel polarization effect. For an $(N,k)$ polar code of $k$ information bits and $N$ encoded bits (\(N = {2^n}\)), an invertible matrix $G_N$ is introduced to describe channel polarization. Here, \({G_N} = G_2^{ \otimes n}\) is a \({2^n} \times {2^n}\) matrix where \({G_2} = \left[ {\begin{array}{*{20}{c}}
   1 & 0  \\
   1 & 1  \\
\end{array}} \right]\)
and \( \otimes n\) denotes the Kronecker product. Let \(u_1^N = ({u_1},{u_2},...,{u_N})\) and \(x_1^N = ({x_1},{x_2},...,{x_N})\) denote the vector of input bits and encoded bits correspondingly, while \(y_1^N = ({y_1},{y_2},...,{y_N})\) denotes the vector of channel output. For the vector channel of copies of a given B-DMC $W(y_i|x_i)$, the transition probability \({W_N}(y_1^N|u_1^N)\) between $u_1^N$ and $y_1^N$ is defined as
\begin{equation}
{W_N}(y_1^N|u_1^N) \buildrel \Delta \over = \prod\limits_{i = 1}^N {W({y_i}|{x_i})}  = \prod\limits_{i = 1}^N {W({y_i}|x_i={{(u_1^NG_N)}_i})}
\end{equation}
and the subchannel \(W_N^{(u_i)}\) with input $u_i$ and output $(y_1^N,u_1^{i-1})$ has the transition probability
\begin{equation}
{W_N^{(u_i)}}(y_1^N,u_1^{i - 1}|{u_i}) = \frac{1}{{{2^{N - 1}}}}\sum\limits_{u_{i + 1}^N} {{W_N}(y_1^N|u_1^N)}.
\end{equation}

The polar encoding scheme is to transmit the set $A$ of $k$ information bits $A=u_1^k$ over $k$ most reliable subchannels out of $N$ subchannels and to use the other ones to transmit the so called ``frozen" bits. Note that \(A = u_1^k\) is labeled with respect of the sequence that information bits are decoded, i.e. $u_1$ is the first decoded information bit while $u_k$ is the last decoded information bit. This labelling is also used in $A_1$ and $A_2$ which will be introduced later. In \cite{Arikan}, SC decoding is used to recover information bits $u_1^k$, where the estimation of information bits \(\widehat{{u_i}}, i \in (1,2,...,k)\), are successively generated by computing the likelihood ratios (LRs) $LR_i$:
\begin{equation}
{LR_i} = \frac{{W_N^{(u_i)} (y_1^N,\widehat{u}_1^{i - 1}|{u_i} = 0)}}{{W_N^{(u_i)} (y_1^N,\widehat{u}_1^{i - 1}|{u_i} = 1)}}.
\end{equation}

LSC decoding is the extension of SC decoding, which is actually a breadth-first searching on the code tree with searching width $L$. LSC decoding keeps a list of size $L$ and updates the list after each \({\widehat{u_i}},i \in (1,k)\), is obtained. It is well known that the complexity of LSC decoding is \(O(LN\log (N))\) with the so called ``lazy copy" strategy, which is $L$ times of the complexity of SC decoding \cite{Ido}  \cite{Chen}. For more about LSC decoding, we refer readers to \cite{Ido} \cite{Chen}.
\section{preliminaries of the proposed algorithm}
LSC decoding with large list size $L$ performs significantly better than SC decoding, but its complexity is also much higher, which is a deficiency in practical implementation.  In this paper, we are interested in finding a low complexity decoding algorithm that can greatly reduce the complexity of LSC decoding while trying to maintain the FER performance of LSC decoding at the same time.

For convenience, we give the definition below before introducing our proposed algorithm.
\begin{definition}
Based on channel polarization, a subchannel $W_N^{({u_i})}$ transmitting information bit $u_i$ is called a ``good" subchannel if and only if its Bhattacharyya parameter $Z(W_N^{({u_i})})$ and all the Bhattacharyya parameters of its subsequential decoded information bits $Z(W_N^{({u_j})}), j \in (i+1,k)$ are smaller than a threshold (i.e. the Bhattacharyya parameter threshold will be discussed later). Otherwise, the subchannel is called a ``bad" one.
\end{definition}

Then we introduce two parameters of the proposed decoding algorithm. One is called ``LR threshold" $LR_{th}$, which determines whether an information bit is received reliably or not. This parameter is used to decide whether to process SC decoding or LSC decoding over $u_i$ based on the observation of $LR_i$. The other parameter is the ``Bhattacharyya parameter threshold" $Z_{th}$. Based on the channel polarization effect, the set of information bits $A$ can be divided into two subsets, namely $A_1=u_1^a$, which stands for the set of information bits that are transmitted over ``bad" subchannels thus are more probable to contribute to FER, and $A_2=u_{a+1}^k$, which stands for the set of information bits that are all transmitted over ``good" subchannels thus are more reliable, given correctly estimated \(u_1^{a}\). The Bhattacharyya parameter threshold $Z_{th}$ is used to decide the number of information bits transmitted over ``bad" subchannels, which is denoted by $a$, and $1 \le a \le k$.

The motivation of the proposed decoding algorithm is to reduce the LSC decoding complexity. We achieve this by two folds. For each information bit, when its estimation is reliable, we process SC decoding rather than LSC decoding. Secondly, when all information bits from ``bad" subchannels are received reliably, we process SC decoding instead of LSC decoding for information bits from ``good" subchannels as well.

Now we briefly describe the proposed LCLSC decoding, which is shown as Algorithm 1. After starting the decoding process, we observe the LRs of information bits in \(A_1\) bit by bit. If the observed LR of $u_i \in A_1$ is larger than the LR threshold $LR_{th}$, SC decoding is processed over $u_i$. If $LR_i$ is greater than $LR_{th}$ for all information bits of subset $A_1$ (i.e. \( LR_i > LR_{th}, i=1,...,a\)), we process SC decoding for the rest of information bits $u_{a+1}^k$. On the other hand, if $LR_i$ is less than the LR threshold $LR_{th}$ for any \(i \in (1,a)\), we process LSC decoding for this information bit and the remaining information bits $u_{i+1}^k$ in $A$.
\begin{algorithm}
\caption{low complexity list successive cancellation decoding}
\begin{algorithmic}[1]

\STATE $a \Leftarrow$  the number of information bits in $A_1$
\STATE $counter=0 $

\FOR {$i=1,2....,a$}
\IF {$LR_{th}$ is satisfied (\( LR_i > LR_{th}\))}
\STATE  process SC decoding over $u_i$
\STATE  $counter$++
\ELSE
\STATE  process LSC decoding over $u_i$
\STATE  break

// \it{This means SC decoding is not applicable after LSC decoding is processed.}

\ENDIF
\ENDFOR

\IF{$counter==a$}
\STATE process SC decoding for the remaining bits
\ELSE
\STATE process LSC decoding for the remaining bits

\ENDIF

\end{algorithmic}
\end{algorithm}

It is clear that in the proposed decoding algorithm, we have two thresholds to determine. One is the LR threshold $LR_{th}$, the other is the Bhattacharyya parameter threshold $Z_{th}$. We discuss these two thresholds in the following sections.

\section{determining the LR threshold}

As discussed above, if the LR of any information bit \({u_i}\in {A_1}\) at the receiver is sufficiently high, i.e. larger than the LR threshold \(LR_{th}\), each of the information bit over ``bad" subchannels is received reliably. Then we only need to process SC decoding instead of LSC decoding since the received signal is more likely to be decoded correctly. For the purpose of determining \(LR_{th}\), we introduce the Proposition below.
\begin{proposition}
In a B-DMC with Bhattacharyya parameter \(Z(W_N^{(i)})\), the upper bound of the bit error probability $Pe(W_N^{(i)})$ in estimating the channel input on the basis of the channel output via the maximun-likehood (ML) decoding is given as follows \cite{Hassani}
\[\frac{1}{2}(1 - \sqrt {1 - Z{{(W_N^{({i})})}^2}} )  \le   Pe(W_N^{(i)}) \le \frac{1}{2} \times Z(W_N^{(i)}).\]
\end{proposition}

Based on Proposition 1, it can be concluded that the lower bound of the probability that the input bit $u_i$ is correctly estimated is \(\left( {1 - \frac{1}{2}Z(W_N^{({u_i})})} \right)\), where $Z(W_N^{(u_i)})$ denotes the Bhattacharyya parameter of the subchannel where $u_i$ is transmitted.

For an input bit $u_{i}$, if the probability of determining $u_i$ as 0 or 1 is smaller than \(\left( {1 - \frac{1}{2}Z(W_N^{({u_i})})} \right)\), we regard $\widehat{u}_i$ is not reliable and need to process LSC decoding over $u_i$. On the other hand, if the probability of determining $u_i$ as 0 or 1 is larger than \(\left( {1 - \frac{1}{2}Z(W_N^{({u_i})})} \right)\), we consider the estimation $\widehat{u_i}$ is reliable and thus employ SC decoding. Therefore, we derive the inequalities which are satisfied when the estimation $\widehat{u}_i$ is reliable
\begin{equation}\label{widehatp}
\left\{ \begin{array}{l}
 \frac{{{W_N^{(u_i)}}(\left. {y_1^N,\widehat{u}_1^{i - 1}} \right|{u_i} = 0)}}{{{W_N^{(u_i)}}(\left. {y_1^N,\widehat{u}_1^{i - 1}} \right|{u_i} = 0) + {W_N^{(u_i)}}(\left. {y_1^N,\widehat{u}_1^{i - 1}} \right|{u_i} = 1)}} >p_i \\
 \;or \\
 \frac{{{W_N^{(u_i)}}(\left. {y_1^N,\widehat{u}_1^{i - 1}} \right|{u_i} = 1)}}{{{W_N^{(u_i)}}(\left. {y_1^N,\widehat{u}_1^{i - 1}} \right|{u_i} = 0) + {W_N^{(u_i)}}(\left. {y_1^N,\widehat{u}_1^{i - 1}} \right|{u_i} = 1)}} >p_i \\
 \end{array} \right.
\end{equation}
and \begin{equation}
p_i= 1 - \frac{1}{2}Z(W_N^{(u_i)})
\end{equation}
which represents the lower bound of correct decoding probability for information bit $u_i$.

When either of the inequalities in (\ref{widehatp}) is satisfied, we process SC decoding instead of LSC decoding over $u_i$. Thus, $LR_{th}$ is defined as

\begin{equation}\label{p}
LR_{th} = \left\{ \begin{array}{l}
 \frac{{p_i}}{{1 - p_i}}\;\;\;\;{LR_i} > 1 \\
 \frac{{1 - p_i}}{{p_i}}\;\;\;\;{LR_i} < 1 \\
 \end{array} \right.
\end{equation}
which means when the observed LR is larger than 1, we process SC decoding instead of LSC decoding if the observed LR is larger than \(\frac{{p_i}}{{1 - p_i}}\;\). When the observed LR is smaller than 1, we do the same if the observed LR is smaller than \(\frac{{1-p_i}}{{p_i}}\;\).

\section{determining the Bhattacharyya parameter threshold }

In this section the Bhattacharyya parameter threshold ${Z_{th}}$ that determines $a$ is derived. We look deeper and exploit the reliability of good polarized subchannels. In consistent with [1], the Bhattacharyya parameter is utilized to measure the reliability of subchannels. Based on Definition 1, ${Z_{th}}$ can be expressed as:
\setcounter{equation}{6}
\begin{equation}\label{Z_th}
\left\{ \begin{array}{l}
 Z(W_N^{({u_i})}) \le {Z_{th}}\:\:\:\:\: \forall i \in u_{a + 1}^k
 \\
 Z(W_N^{({u_a})}) > {Z_{th}}. \\
 \end{array} \right.
\end{equation}

Since the bit error events in SC decoding are not independent, the FER lower bound of the ML decoding \(P_e^{ML}\) is derived according to Proposition 1
\begin{align}
P_e^{ML} &\ge 1 - \prod\limits_{i \in A} {\left( {1 - Pe(W_N^{({u_i})})} \right)} {\rm{ }} \nonumber  \\ & \ge  1 - \prod\limits_{i \in A} {\left( {1 - \frac{1}{2}(1 - \sqrt {1 - Z{{(W_N^{({u_i})})}^2}} )} \right)}
\end{align}

In [1], $\sum\limits_{i \in A}^{} {Z(W_N^{({u_i})})}$  serves as the FER upper bound of the SC decoding \(P_e^{sc}\). Thus, we have
\begin{align}
\sum\limits_{i \in A}^{} {Z(W_N^{({u_i})})}  \ge 1 - \prod\limits_{i \in A} {} \left( {1 - \frac{1}{2}(1 - \sqrt {1 - Z{{(W_N^{({u_i})})}^2}} )} \right)
\end{align}

It is noted that $\sum\limits_{i \in A}^{} {Z(W_N^{({u_i})})}$ consists of the Bhattacharyya parameters of $k$ different subchannels transmitting information bits. Some of them are quite small leading to reliable subchannels, others are large resulting in unreliable subchannels.  As FER mainly results from subchannels with larger \({Z(W_N^{(u_i)})}\), we could determine the Bhattacharyya parameter threshold as the one that can approach the FER lower bound of ML decoding :
\begin{equation}
k \times {Z_{th}} = 1 - \prod\limits_{i \in A} {} \left( {1 - \frac{1}{2}(1 - \sqrt {1 - Z{{(W_N^{({u_i})})}^2}} )} \right)
\end{equation}
which can be explained as follows. We consider a subchannel \({W_N^{(u_i)}}\) where $u_i$ is transmitted. If \({Z(W_N^{(u_i)})}\) is larger than $Z_{th}$, then we have \(k \times {Z(W_N^{({u_i})})} \) larger than the FER lower bound of ML decoding. Thus we consider the subchannel \({W_N^{(u_i)}}\) less reliable. Therefore, to achieve a good FER performance, we should observe whether the estimation $\widehat{u_i}$ satisfies $LR_i$. If $LR_i$ is satisfied, we regard $u_i$ is reliably recovered. Otherwise, we should process LSC decoding over $u_i$ to approach the FER performance of ML decoding (note LSC decoding becomes ML decoding when $L=2^k$, and practically the performance of LSC decoding is very close to ML decoding with moderate $L$). If \({Z(W_N^{(u_i)})}\) is less than $Z_{th}$, we have \(k \times {Z(W_N^{({u_i})})}\) lower than the FER lower bound of $P_e^{ML}$. Therefore, \({W_N^{(u_i)}}\) is considered a more reliable subchannel and then the SC decoding is likely to provide correct estimation of the information bit, if the LR thresholds of $u_1^a$ are all satisfied. As mentioned above, the information bits in $u_{a+1}^k$ all have a Bhattacharyya parameter smaller than $Z_{th}$. Therefore, it is reasonable to process SC decoding over $u_{a+1}^k$ to approach the FER of ML decoding if the estimation of $u_1^a$ is reliable.

Based on the discussion above, the Bhattacharyya parameter threshold $Z_{th}$ is determined as
\begin{equation}
{Z_{th}} = \frac{1}{k} \times \left\{ {1 - \prod\limits_{i \in A} {} \left( {1 - \frac{1}{2}(1 - \sqrt {1 - Z{{(W_N^{({u_i})})}^2}} } \right)} \right\}
\end{equation}
and $a$ can be obtained according to (\ref{Z_th}), as $W_N^{(u_{a+1})}$ is the first ``good" subchannel in the decoding process.


\section{complexity analysis and simulation results}

%

In this section we first analyze the complexity of the LCLSC decoding. Note in the LCLSC decoding, the SC decoding is processed over some (or all) information bits while the LSC decoding is processed over the rest ones. Denote \(m\) as the average number of information bits over which we process SC decoding and thus \(k-m\) is the average number of information bits over which the LSC decoding is processed. In consistent with \cite {Arikan}, the computational model for complexity analysis is a single processor machine with random-access memory, and the complexities expressed are time complexities. For the decoding algorithms, the time complexities are measured with the total number of LR calculations. Note that the time complexity of SC and LSC decoding is \({\rm O}(N\log (N))\) and \({\rm O}(LN\log (N))\) respectively, meaning the number of LR calculations is $(N + N\log (N))$ and $L\times(N + N\log (N))$ correspondingly \cite{Arikan}. Then the average number of LR calculations of our proposed LCLSC decoding algorithm is given by
\begin{equation}
\begin{array}{l}
 \;\;\; C = \:\frac{m}{k} \times (N + N\log (N)) + \frac{{k - m}}{k} \times L(N + N\log (N)). \\
 \end{array}
\end{equation}

When implementing the LCLSC decoding algorithm, there may be the case that all the information bits are recovered with SC decoding and there may also be the case that some information bits are recovered with SC while others with LSC decoding. So the complexity in (12) is actually an averaged complexity. Also, the complexity of LCLSC decoding in the Figures below are all averaged over the simulation. It is straightforward that \(C\) is less than $L\times(N + N\log (N))$, thus the proposed LCLSC decoding has a lower decoding complexity than LSC decoding, which will be shown in the following. The saving  in decoding complexity is considerable for low code rates.

Now, we present simulation results of a polar code with length 512 and different code rates. The results for the BEC with erasure rate \(\varepsilon=0.4 \), the BSC with cross probability 0.11 and the BAWGNC with the standard deviation of Gaussian noise \(\sigma  = 0.97865\) are depicted. Figs 1, 2 and 3 show the FER performance of the SC, LSC and LCLSC decoding on various channels. The capacity for both the BSC and the BAWGNC is 0.5 and the codes for BSC and BAWGNC are those optimzed via Arikan's heuristic method [2]. In the LCLSC decoding, the LR threshold is determined by the correct decoding probability for each information bit $p_i$, as discussed above. In the results of the BEC, we set $p_i$ in two different ways: (i) set $p_i$ to be its lower bound \(\left( {1 - \frac{1}{2}Z(W_N^{({u_i})})} \right)\), as (5), and (ii) set $p_i$ to be a fixed value 0.9. As the polarization indices are not known in closed form for the BSC and the BAWGNC, $p_i$ is therefore set to 0.9 in those channels. List size $L$ is set to 16 in both the LSC and LCLSC decoding.

From Figs 1, 2 and 3 we can see that the LCLSC decoding has almost the same FER performance as LSC decoding and much better FER performance than SC decoding. Figs 4, 5 and 6 show the corresponding complexity of the three decoding algorithms on various channels. It is illustrated that the LCLSC decoding has a lower complexity than the LSC decoding. The complexity reduction is larger with lower code rate, as there are more ``reliable" subchannels where we could process SC decoding instead of LSC decoding. Especially, in low to medium code rates, the complexity of LCLSC decoding is near to that of SC decoding with slightly degraded FER performance compared with LSC decoding.

\begin{center}
\includegraphics[height=4.5cm]{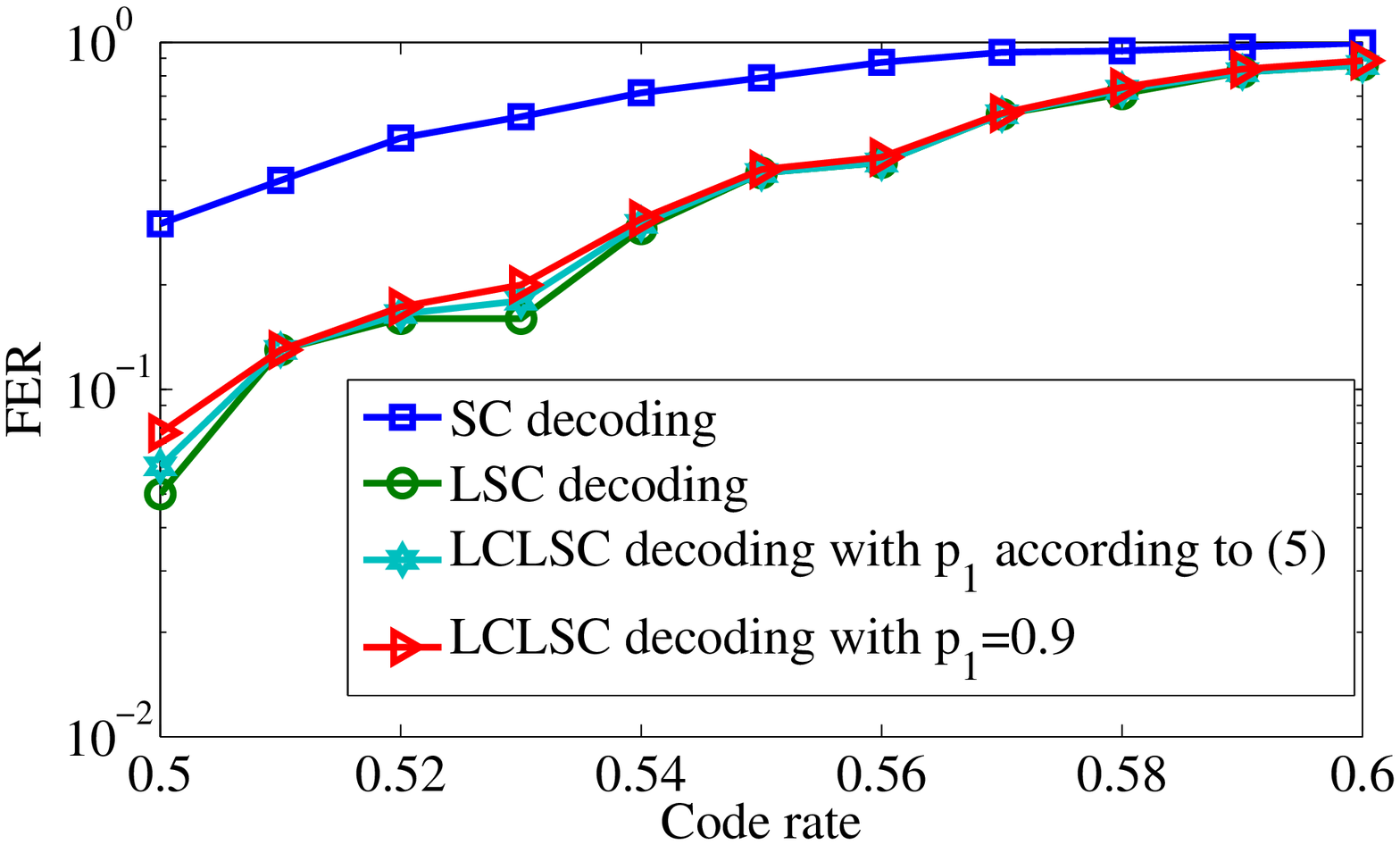}
\makeatletter\def\@captype{figure}\makeatother
\caption{ FER comparison of LSC decoding, SC decoding and LCLSC decoding over the BEC of channel erasure rate \(\varepsilon=0.4 \), $N=512$. }
\end{center}

\begin{center}
\includegraphics[height=4.5cm]{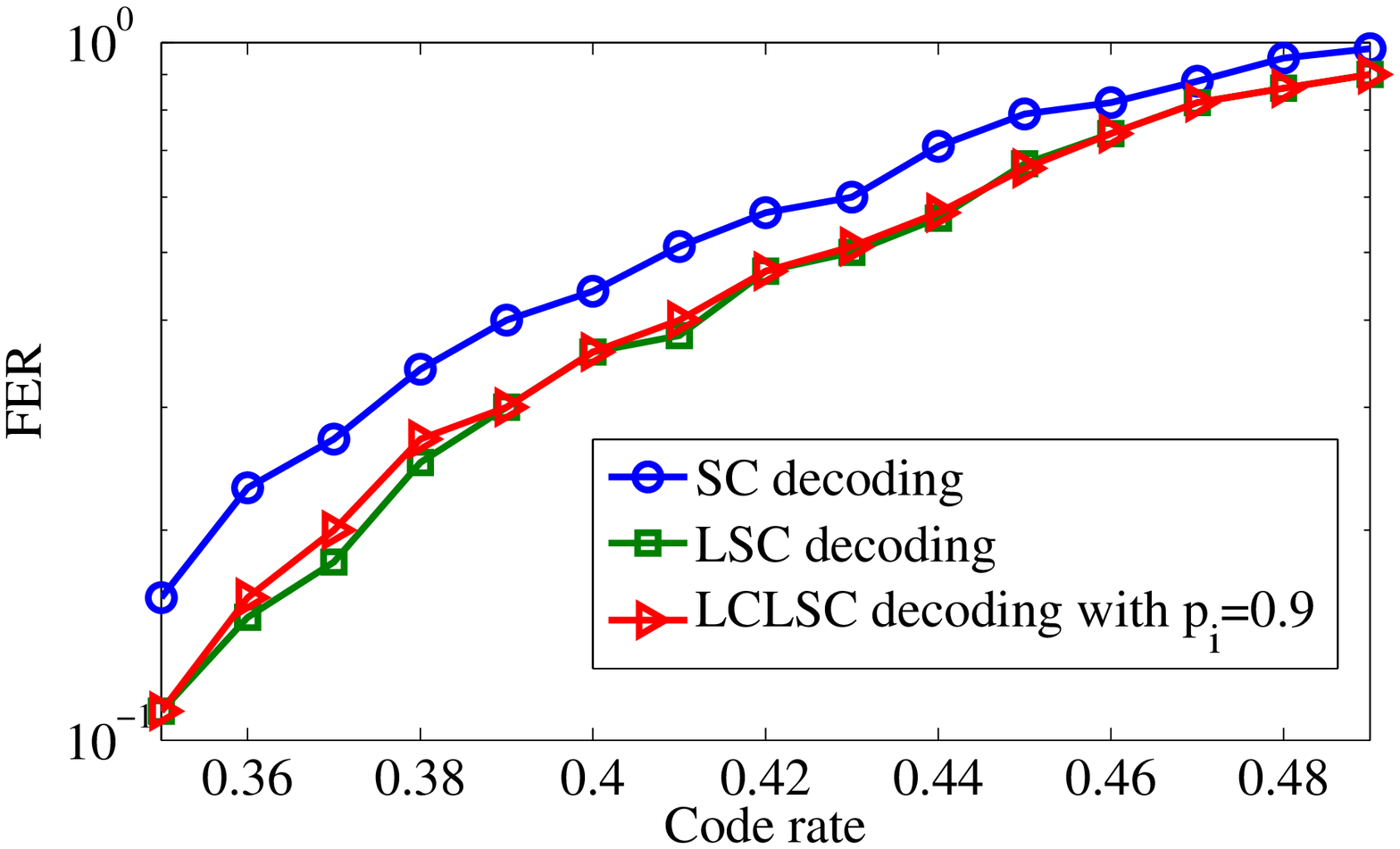}
\makeatletter\def\@captype{figure}\makeatother
\caption{ FER comparison of LSC decoding, SC decoding and LCLSC decoding over the BSC of cross probability 0.11, $N=512$.  }
\end{center}

\begin{center}
\includegraphics[height=4.5cm]{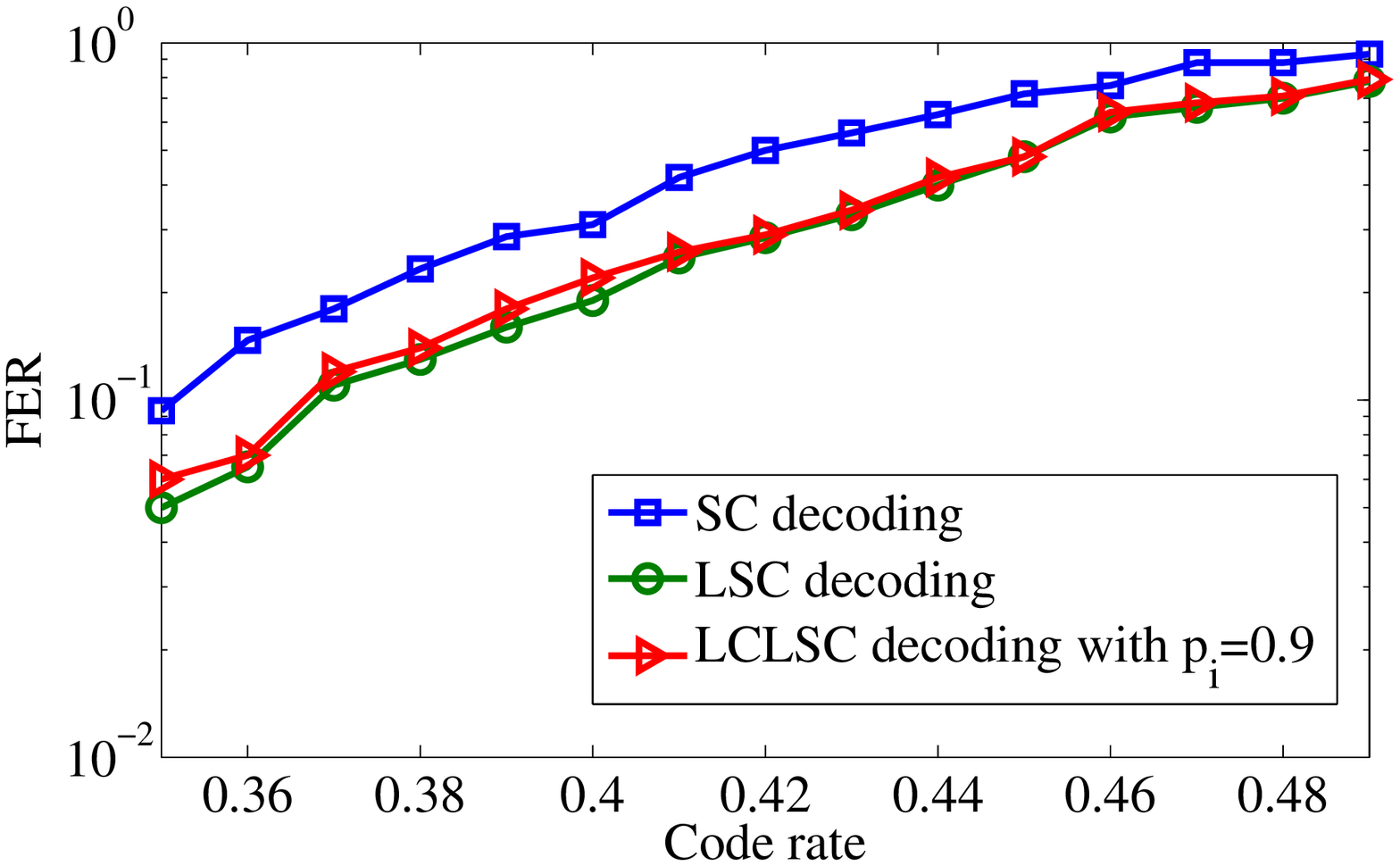}
\makeatletter\def\@captype{figure}\makeatother
\caption{ FER comparison of LSC decoding, SC decoding and LCLSC decoding over the BAWGNC of standard deviation of Gaussian noise \(\sigma  = 0.97865\), $N=512$.}
\end{center}

\begin{center}
\includegraphics[height=4.5cm]{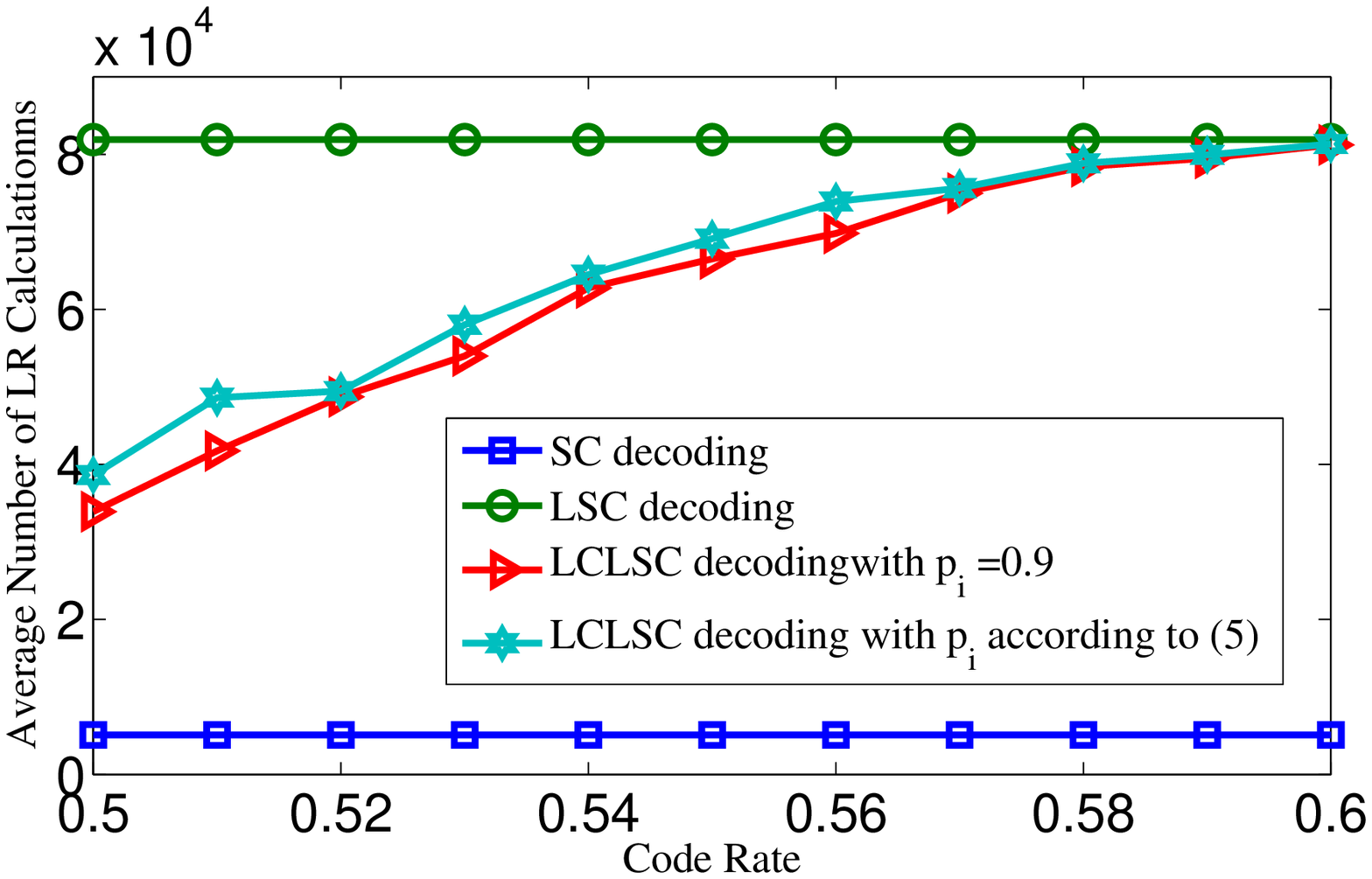}
\makeatletter\def\@captype{figure}\makeatother
\caption{ Complexity comparison of LSC decoding, SC decoding and LCLSC decoding over the BEC of channel erasure rate  \(\varepsilon=0.4 \), $N=512$.}
\end{center}

\begin{center}
\includegraphics[height=4.5cm]{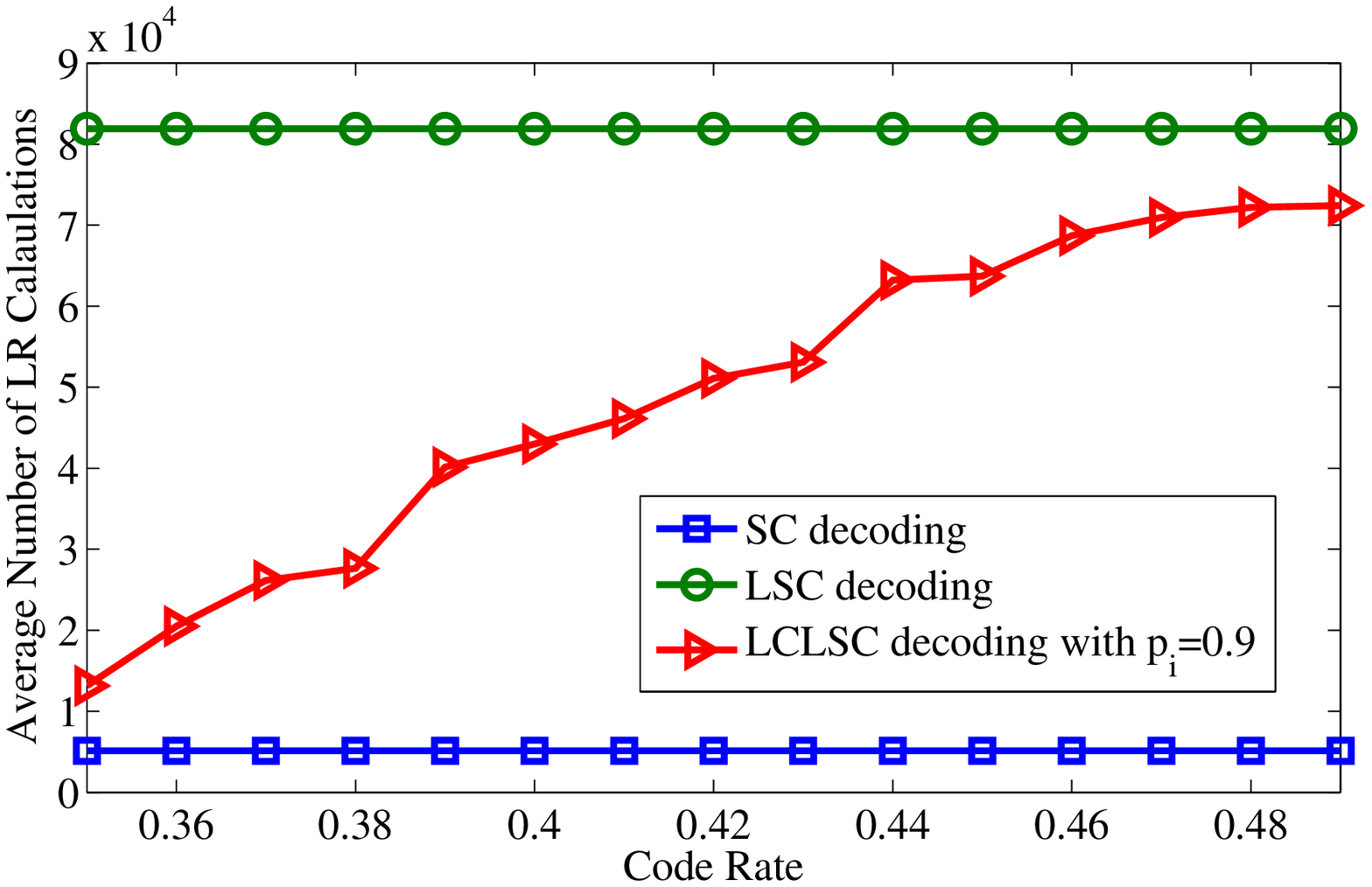}
\makeatletter\def\@captype{figure}\makeatother
\caption{ Complexity comparison of LSC decoding, SC decoding and LCLSC decoding over the BSC of cross probability 0.11, $N=512$.}
\end{center}

\begin{center}
\includegraphics[height=4.5cm]{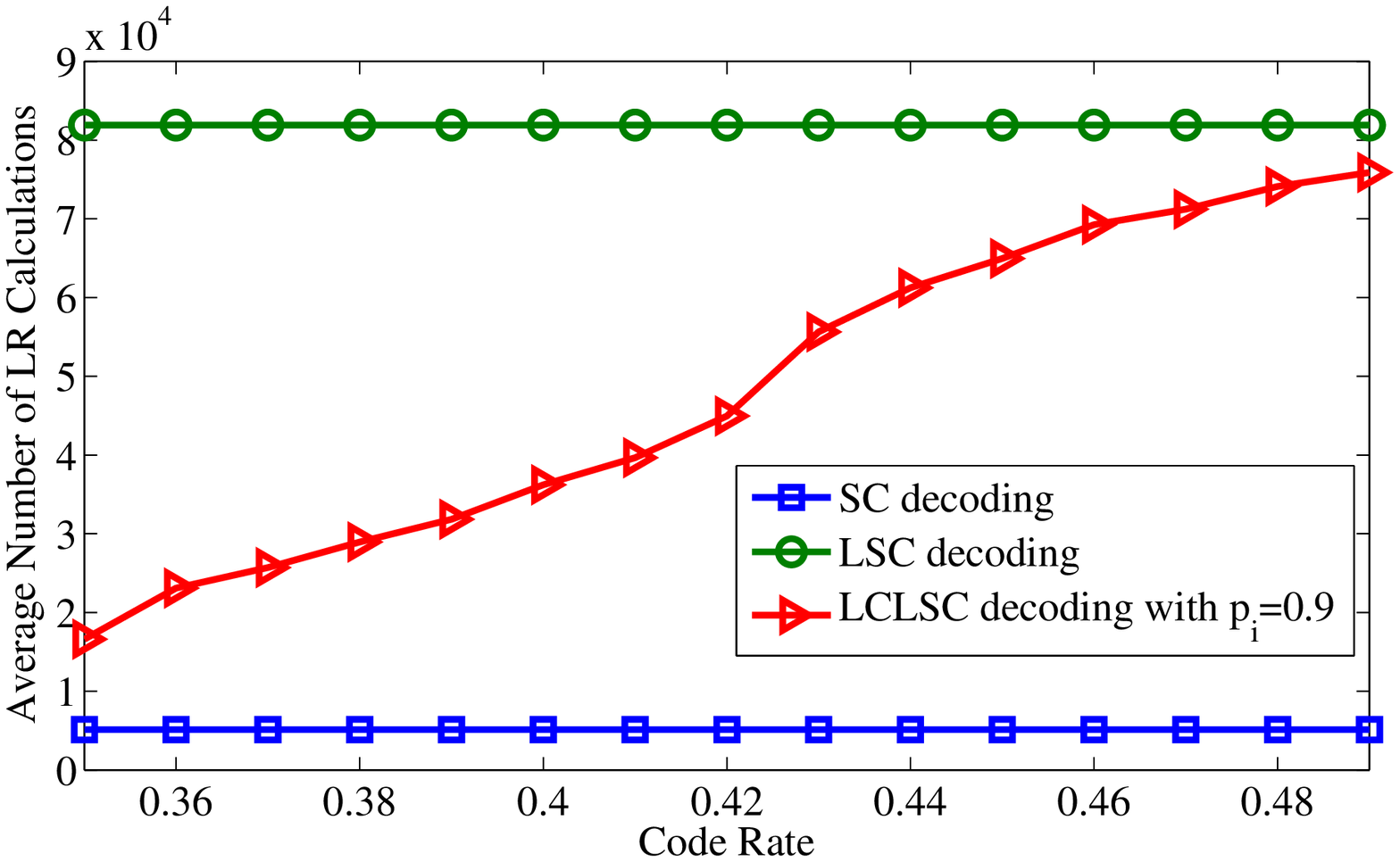}
\makeatletter\def\@captype{figure}\makeatother
\caption{ Complexity comparison of LSC decoding, SC decoding and LCLSC decoding over the BAWGNC of standard deviation of Gaussian noise \(\sigma  = 0.97865\), $N=512$.}
\end{center}

\section{conclusion}
In this paper, an LCLSC decoding algorithm that can reduce the complexity of LSC decoding was proposed. We set an LR threshold and a Bhattacharyya parameter threshold to determine the information bits over which SC decoding instead of LSC decoding could be utilized. Simulation results showed that the proposed decoding algorithm could reduce the decoding complexity of LSC decoding significantly with low code rate while almost maintaining the same FER performance.

\end{document}